\documentclass[11pt]{article}
\usepackage{moriond,epsfig}

\def\Journal#1#2#3#4{{#1} {\bf #2}, #3 (#4)}

\def\NIMA{{\em Nucl. Instrum. Methods} A}
\def\NPB{{\em Nucl. Phys.} B}
\def\PLB{{\em Phys. Lett.}  B}
\def\PRL{\em Phys. Rev. Lett.}
\def\PRD{{\em Phys. Rev.} D}

\def\JHEP{\em JHEP} 

\def\mco{\multicolumn}

\def\be{\begin{equation}}
\def\ee{\end{equation}}
\def\bea{\begin{eqnarray}}
\def\eea{\end{eqnarray}}

\begin{document}
\title{MEASUREMENTS OF B RARE DECAYS AT THE TEVATRON}

\author{ F. SCURI \\
on behalf of the CDF and D0 collaborations.
}

\address{Istituto Nazionale di Fisica Nucleare, Sezione di Pisa, \\ 
Largo B. Pontecorvo 3, I-56127 Pisa, Italy}

\maketitle\abstracts{
A summary of recent results on B rare decays from the CDF and D0 experiments
operating in Run II of the Fermilab Tevatron is given; 
analyzed decay modes are B$_{d,s}
\rightarrow$ hh, B$_{d,s} \rightarrow \mu^{+} \mu^{-}$, and B $\rightarrow 
\mu^{+} \mu^{-}$ h. Data samples are relative to 
1 fb$^{-1}$ or more integrated luminosity of $p\bar{p}$ collisions at
$\sqrt{s}$ = 1.96 TeV. 
All reported results are in agreement with Standard Model predictions and
consistent with B-Factories analyzes. 
}

\section{Introduction}

The large production of all kinds of $b$-hadrons at the Tevatron offers the 
opportunity to study rare decays also in the B$_{s}$ and $b$-baryon sectors,
exploiting a physics program complementary to the B-Factories.

The bottom anti-bottom production cross-section $\sigma(b\bar{b})$ at the
Tevatron is O(10$^{5}$) larger than production in $e^{+}e^{-}$ colliders at
the $\Upsilon(4s)$ or $Z^{0}$ energy scale; however, the inelastic 
cross-section is a factor 10$^{3}$ larger than $\sigma(b\bar{b})$ and the
branching ratios of rare $b$-hadron decays are O(10$^{-5}$) or lower;
therefore, interesting events must be extracted from a high track multiplicity
environment and detectors need to have very good tracking and vertex 
resolution, wide acceptance and good particle identification, highly 
selective trigger. Both CDF and D0 detectors, whose detailed description can be
found elsewhere \cite{det}, match those requirements. 

The following decays modes are analyzed here: charmless B$_{d,s}
\rightarrow h^{+}h{-}$, B$_{d,s} \rightarrow \mu^{+} \mu^{-}$, and B $\rightarrow 
\mu^{+} \mu^{-}$ h.
All those rare decay modes are interesting in the search of New Physics 
contributions, each mode with its own peculiarity: charmless B
$\rightarrow$ hh decays are a useful tool for probing CKM inferring on 
the $\alpha$ and $\gamma$ angles of the Unitarity Triangle; they are also 
sensitive to New Physics effects both in the Penguin diagram contributions 
to the process and via anomalies in the CP Asymmetry $A_{CP}$.
 
The Standard Model prediction of the Branching Ratio for flavor-changing 
neutral current (FCNC) processes like B$_{s} \rightarrow 
\mu^{+} \mu^{-}$ is very suppressed: $\mathcal{B}(B_{s} \rightarrow 
\mu^{+} \mu^{-})^{SM}$ = (3.35$\pm 0.32) \times 10^{-9}$ (M.Blanke {\it et al.}
\cite{Bla}); 
a slightly higher value for the BR upper limit was recently 
estimated in Constrained Minimal Flavor Violation \cite{Hur}: 
$\mathcal{B}(B_{s} 
\rightarrow \mu^{+} \mu^{-})^{CMFV} < ~7.42 \times 10^{-9} ~(95\% ~C.L.)$; 
the leptonic branching fraction of the B$_{d}$ decay is suppressed by a factor
$\mid V_{td}/V_{ts} \mid ^{2}$ in the CKM matrix elements leading to a SM
predicted BR of O(10$^{-10}$). The decay amplitude of B$_{d,s} 
\rightarrow \mu^{+} \mu^{-}$ can be significantly enhanced in some extensions
of the SM, such as, for instance, the type-II two-Higgs-doublet-model (2HDM),
where the BR depends only on M$_{H^{+}}$ and on tan$\beta$, the ratio between
the vacuum expectation values of the two neutral Higgs fields; in this 
case \cite{Log}, $\mathcal{B}
(B \rightarrow \mu^{+} \mu^{-})^{2HDM} \propto$ (tan$\beta)^{4}$. In the 
minimal super-symmetric standard model (MSSM) the dependence on tan$\beta$ is
even stronger, 
$\mathcal{B}(B \rightarrow \mu^{+} \mu^{-})^{MSSM} 
\propto$ (tan$\beta)^{6}$, leading to an enhancement of orders of magnitude
\cite{Bab} in case of large values of tan$\beta$.

The $b \rightarrow s ~\mu^{+}\mu^{-}$ transition in the B$\rightarrow$ 
$s$-meson
$\mu^{+}\mu^{-}$ modes allows the study of FCNC in more detail through
additional observables, such as the dimuon invariant mass m($\mu^{+}\mu^{-}$),
and the forward-backward asymmetry of the $s$-quark in the dimuon system.

A summary of recent results from the Tevatron experiments is given in
the following section for all modes; quoted values refer
to integrated luminosities varying from about 1 fb$^{-1}$ to about 2
fb$^{-1}$, depending on the specific analysis, and covering the Tevatron Run-II
data taking period from 2002 to beginning of 2007.

\section{Tevatron result summary for B rare decays} 

Hadronic decays of the B mesons into two charged tracks are studied in an 
efficient way by the CDF experiment, exploiting the high performances of
its peculiar Secondary Vertex Trigger (SVT), designed to select events
with vertexes displaced with respect to the primary vertex and whose 
full description can be found elsewhere \cite{SVT}. Details of the event
selection procedure and of the analysis method can be found in a recent
CDF publication \cite{CDF1}.

\begin{figure}[htb]
$\begin{array}{lll}
\psfig{figure=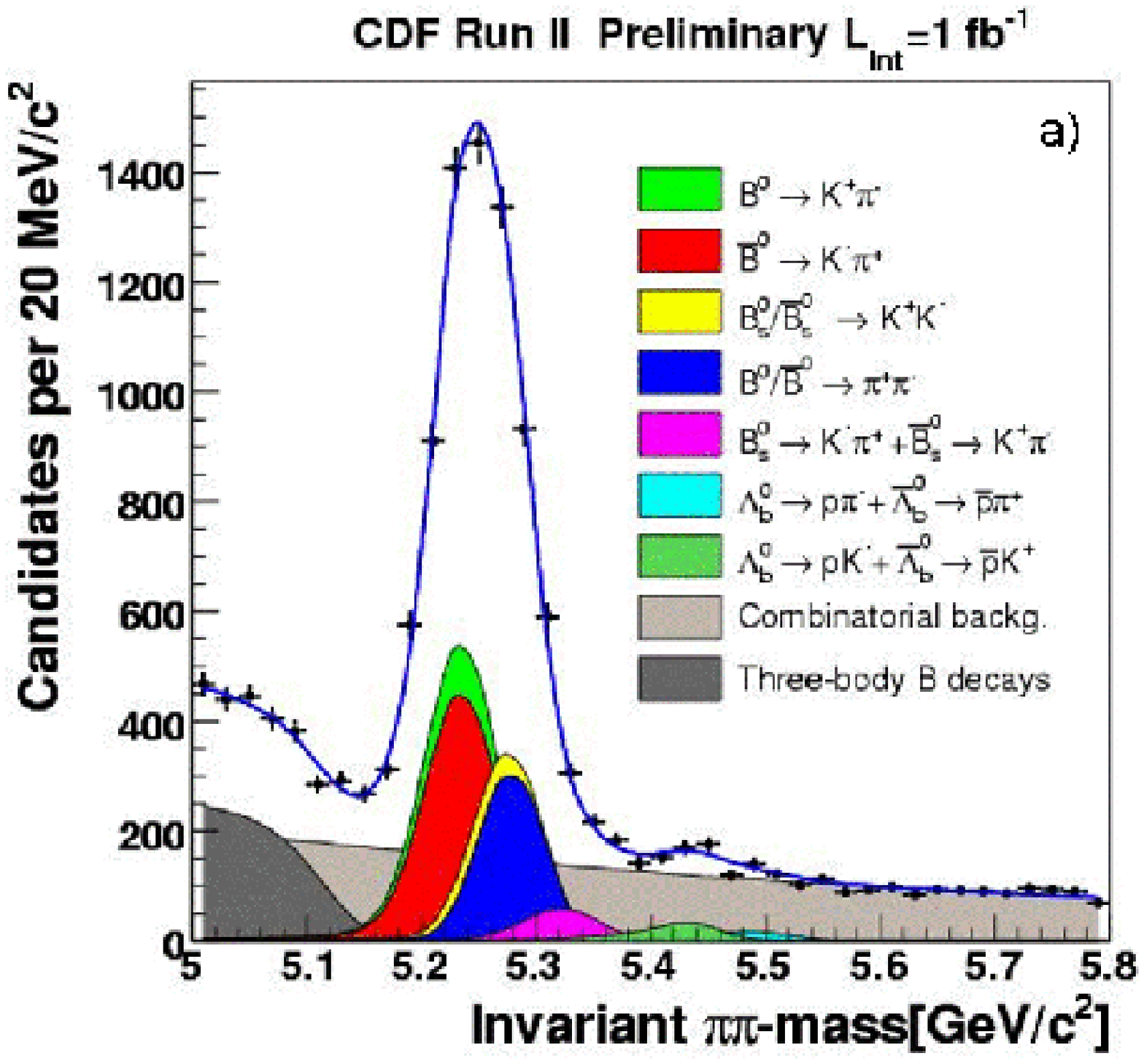,height=4.cm} & 
\psfig{figure=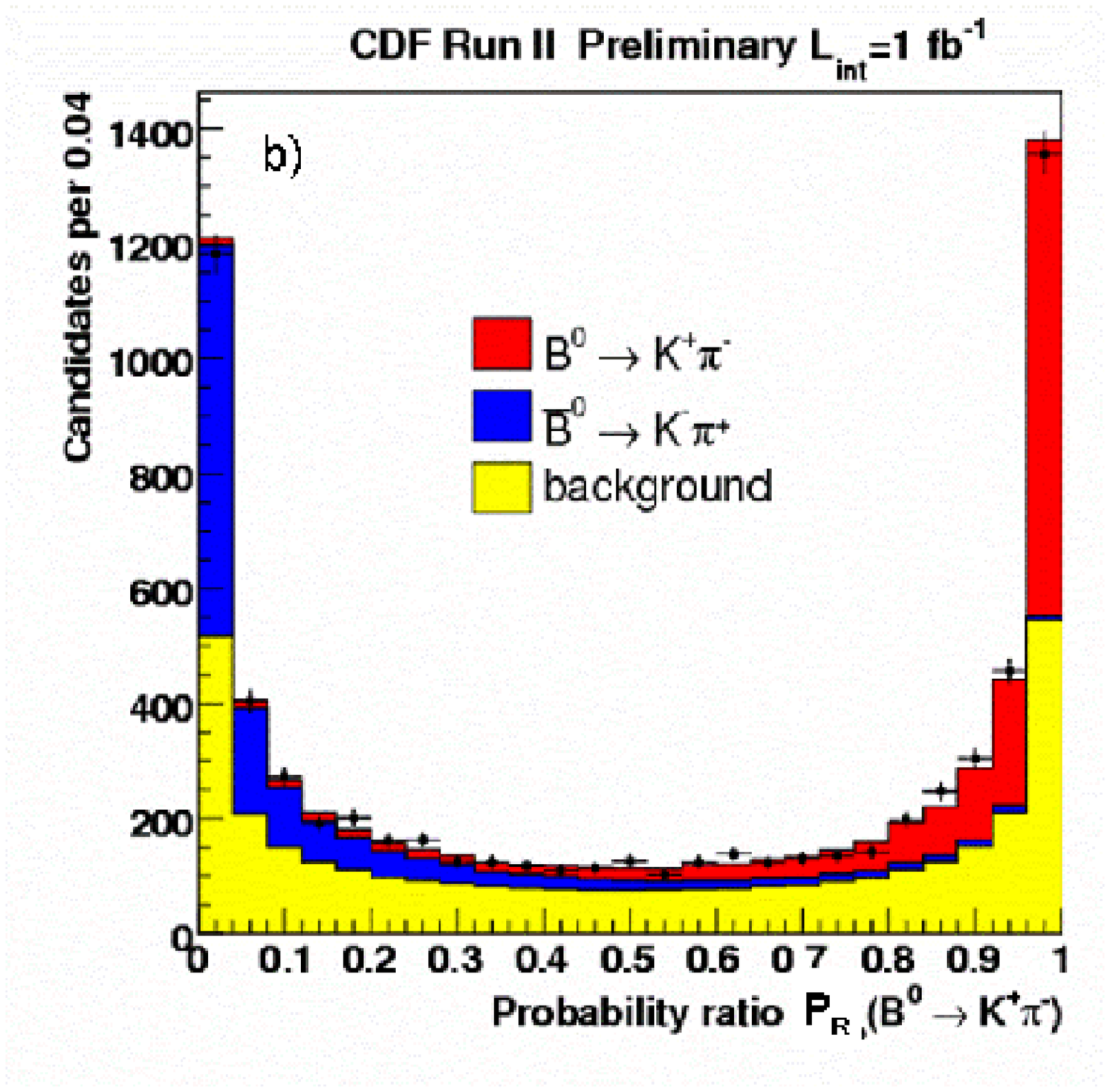,height=4.cm} &
\psfig{figure=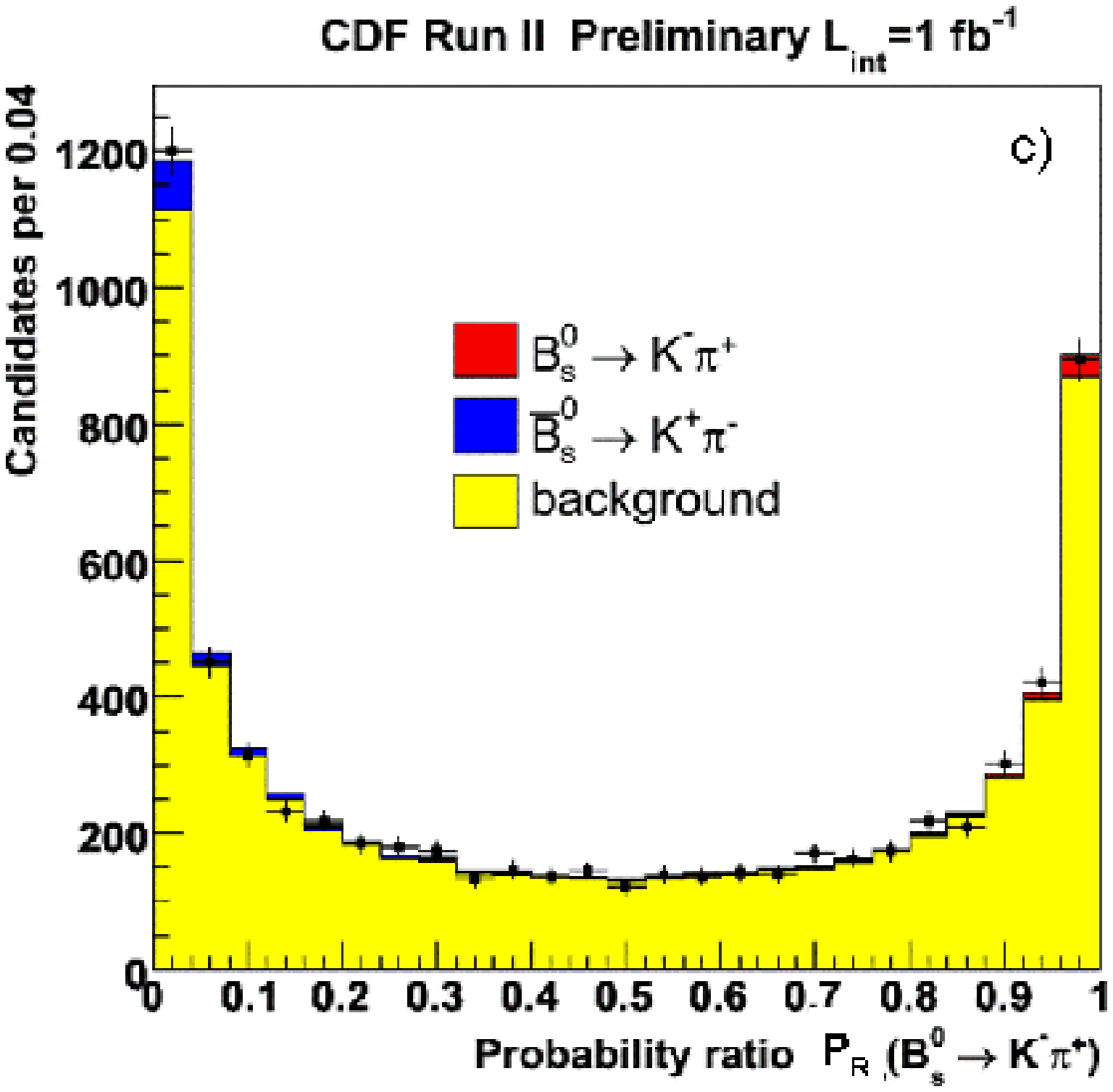,height=4.cm}
\end{array}$
\caption{a): M($\pi \pi$) invariant mass distributions; b) Probability
Ratio for B$^{0}\rightarrow K^{+}\pi^{-}$; c) Probability ratio for 
B$_{s}\rightarrow K^{-}\pi^{+}$.
\label{fig:fig1}}
\end{figure}

The invariant mass m($\pi \pi$) after cut optimization is shown in figure 1a);
about 6500 signal events accumulate in the signal region; individual modes
overlap in a single peak as shown by the signal composition (filled colored
areas) estimated via Monte Carlo and including full detector simulation. 
The signal
composition in real data is measured with a likelihood fit, combining
information from kinematics (mass and momentum) and particle identification;
details of the fit procedure can be found elsewhere \cite{CDF1}.

\begin{table}[htb]
\caption{CDF results with 1 fb$^{-1}$ large sample for $\mathcal{B}(B
\rightarrow h^{+}h^{-})$.
\label{tab:exp}}
\begin{center}
\begin{tabular}{|l|c|}
\hline
Decay mode &
$BR ~\times ~10^{6}$ 
\\ \hline
B$^{0}\rightarrow \pi^{+}\pi^{-}$ & $5.10 ~\pm ~0.33(stat.) ~\pm ~0.36(syst.)$ 
\\
B$^{0}\rightarrow K^{+}K^{-}$ & $0.39 ~\pm ~0.16(stat.) ~\pm ~0.12(syst.)$
\\ \hline
B$_{s}\rightarrow K^{+}K{-}$ & $24.4 ~\pm ~1.4(stat.) ~\pm ~4.6(syst.)$ 
\\
B$_{s}\rightarrow \pi^{+}K^{-}$ & $5.0 ~\pm ~0.75(stat.) ~\pm ~1.0(syst.)$
\\
B$_{s}\rightarrow \pi^{+}\pi^{-}$ & $0.53 ~\pm ~0.31(stat.) ~\pm ~0.40(syst.)$
\\ \hline
\end{tabular}
\end{center}
\end{table}

Measured branching ratios for individual modes are summarized in table 
\ref{tab:exp}; values for the B$_{d}$ modes are consistent with the B-Factories
results, while the B$_{s}$ modes are CDF exclusive.

The very good separation power between $B^{0}$ and $\bar{B^{0}}$ achieved 
with the likelihood fit is shown in figure 1b) for the raw asymmetry 
of the B$^{0}\rightarrow K^{+} \pi^{-}$ mode ; the probability ratio P$_{R} = 
pdf(B^{0})/[pdf(B^{0})+pdf(\bar{B^{0}})]$ is based on probability density
functions (pdf) using 5 observables (M($\pi_{1},\pi_{2}$), $p_{1}+p_{2}$,
the charged pion momentum um-balancing \cite{CDF1}, and the particle
identification ID$_{1}$ and ID$_{2}$). With this method, the first measurement
of direct CP asymmetry (A$_{CP}$) in the B$_{s}$ system was done, figure 1c). 
CDF results on A$_{CP}$ for the $K \pi$ modes are given in the following 
equations:  
\begin{equation}
\begin{array}{lll}
A_{CP}(B^{0}_{d}\rightarrow K^{+}\pi^{-}) & = 
\frac{N(\bar{B^{0}_{d}}\rightarrow K^{-}\pi^{+})- 
N(B^{0}_{d}\rightarrow K^{+}\pi^{-})}
{N(\bar{B^{0}_{d}}\rightarrow K^{-}\pi^{+})
+N(B^{0}_{d}\rightarrow K^{+}\pi^{-})} & 
= -0.086 \pm 0.023(stat.) \pm 0.009(syst.)\\ \\
A_{CP}(B^{0}_{s}\rightarrow K^{-}\pi^{+}) & = 
\frac{N(\bar{B^{0}_{s}}\rightarrow K^{+}\pi^{-})- 
N(B^{0}_{s}\rightarrow K^{-}\pi^{+})}
{N(\bar{B^{0}_{s}}\rightarrow K^{+}\pi^{-})
+N(B^{0}_{s}\rightarrow K^{-}\pi^{+})} & 
= -0.39 \pm 0.15(stat.) \pm 0.08(syst.)
\end{array}\label{eq:eq1}
\end{equation}
Tevatron and B-Factories results on A$_{CP}(B^{0}_{d}\rightarrow 
K^{+}\pi^{-})$ agree; Tevatron confirms the difference in sign w.r.t. 
A$_{CP}(B^{+}\rightarrow K^{+}\pi^{0})$. The measured value of
A$_{CP}(B^{0}_{s}\rightarrow K^{-}\pi^{+})$ (\ref{eq:eq1}) can be compared with the
value obtained with the Lipkin test \cite{Lip}, based on minimal
assumptions, just SM with SU(3) symmetry; in this case, 
the following equation holds:  
\begin{equation}
A_{CP}(B^{0}_{s}\rightarrow K^{-}\pi^{+}) =
- A_{CP}(B^{0}_{d}\rightarrow K^{+}\pi^{-}) \cdot 
\frac{\mathcal{B}(B^{0}_{d}\rightarrow K^{+}\pi^{-})} 
{\mathcal{B}(B^{0}_{s}\rightarrow K^{-}\pi^{+})} \cdot
\frac{\tau(B^{0}_{d})}{\tau(B^{0}_{s})} \approx 0.37
\label{eq:eq2}
\end{equation}
In equation \ref{eq:eq2} the values of A$_{CP}(B^{0}_{d}\rightarrow 
K^{+}\pi^{-})$ and $\mathcal{B}(B^{0}_{d}\rightarrow K^{+}\pi^{-})$ are
taken from the Heavy Flavor Averaging Group \cite{HAF},  
$\mathcal{B}(B^{0}_{d}\rightarrow K^{+}\pi^{-})$ is the CDF resulted quoted
in table \ref{tab:exp}, and the lifetime ratio is set equal to one. 

\begin{figure}[htb]
\begin{center}
$\begin{array}{rr}
\psfig{figure=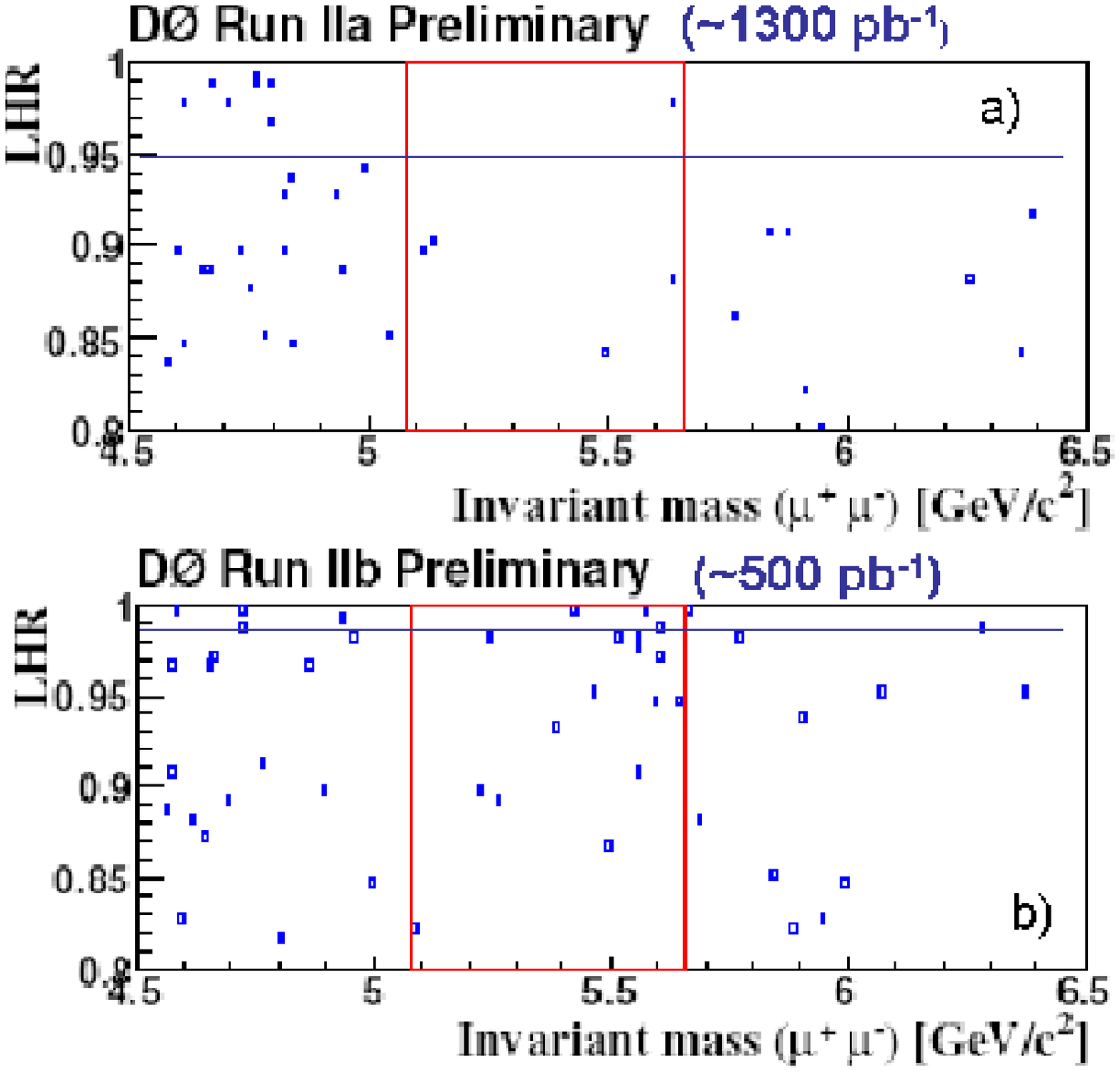,height=4.5cm} & 
\psfig{figure=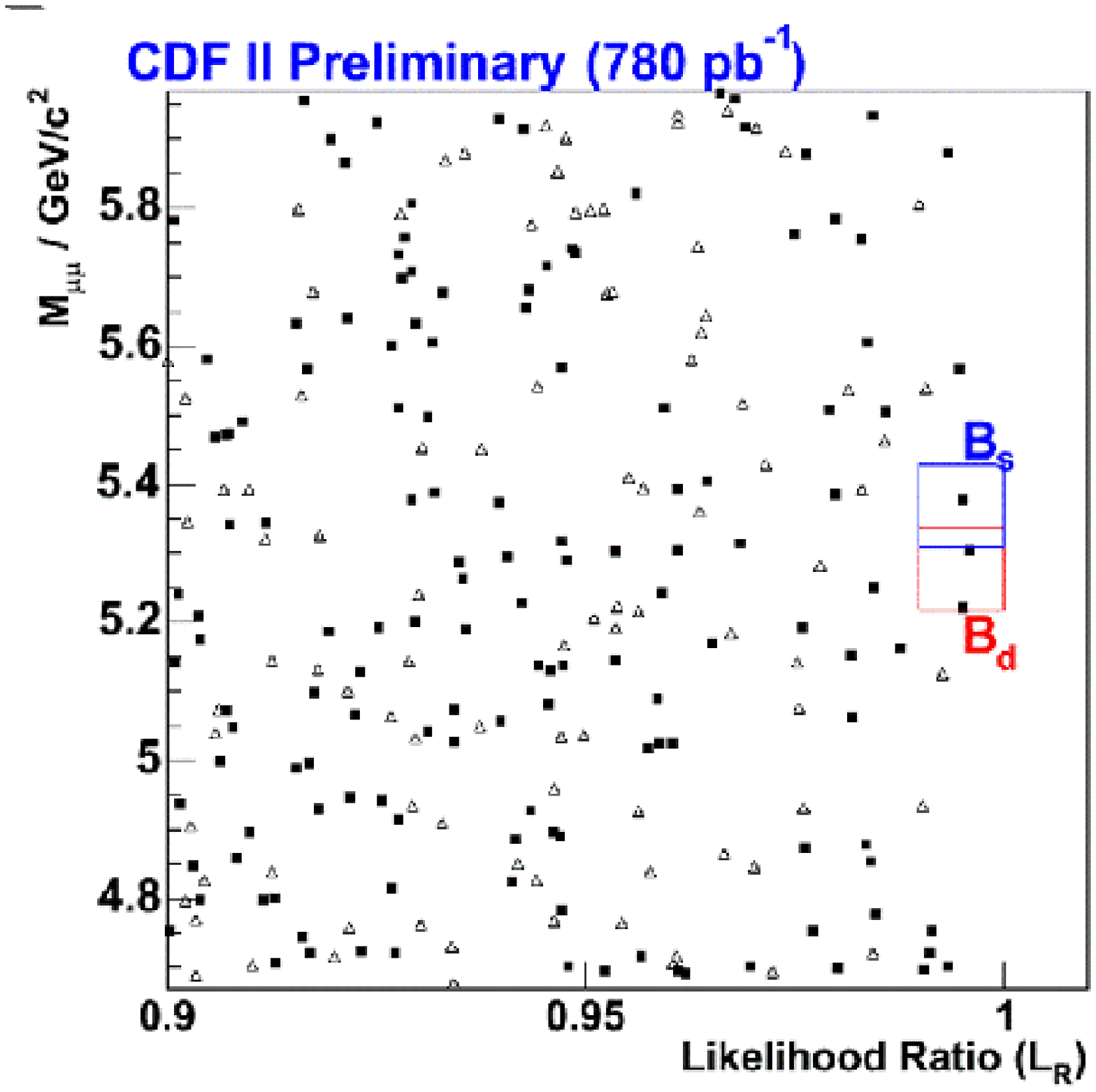,height=4.5cm} 
\end{array}$
\caption{a), b): D0 Likelyhood Ratio L$_{R}$ vs. $\mu - \mu$ invariant mass; 
blue lines set the L$_{R}$ optimal value; c) CDF $\mu - \mu$ invariant mass
vs. L$_{R}$; red and blue boxes define the search areas for B$_{d}$ and
B$_{s}$ respectively.
\label{fig:fig2}}
\end{center}
\end{figure}

The two Tevatron experiments adopt similar procedures to search for
B$_{s}\rightarrow \mu^{+} \mu^{-}$ events; a blind cut optimization
is made using signal Monte Carlo samples and signal sidebands for
estimation of the background in the data sample. The B$^{+}
\rightarrow J/\psi ~K^{+}$ mode is used as normalization channel in the
BR estimation. Similar discriminating 
observables (secondary vertex displacement, B pointing angle to the 
primary vertex, B isolation) are used to construct a Likelihood Ratio
(L$_{R}$); optimal values of L$_{R}$ for event counting are obtained with 
different methods.
The main difference in the two analyzes is the definition of the signal
window in the $\mu - \mu$ invariant mass spectrum (figure \ref{fig:fig2}); 
a 360 MeV/c$^{2}$ wide 
search region including B$_{d}$ and B$_{s}$ is defined by D0, two different
regions, 120 MeV/$^{2}$ wide and centered in the B$_{d}$ and B$_{s}$ nominal 
masses are used by CDF, quoting separate limits; D0 assumes
$\mathcal{B}(B_{d}\rightarrow \mu^{+} \mu^{-}) \ll
\mathcal{B}(B_{s}\rightarrow \mu^{+} \mu^{-})$ and quotes the overall
value as a conservative limit on 
$\mathcal{B}(B_{s}\rightarrow \mu^{+} \mu^{-})$.

Tevatron results on B$_{s}\rightarrow \mu^{+} \mu^{-}$ are summarized
in table \ref{tab:exp2}; a statistical combination is made by D0 to
quote the overall limit from two data taking periods, before (run IIa)
and after (run IIb) insertion of a new inner layer in silicon vertex
detector.  
   
\begin{table}[htb]
\caption{Tevatron result summary for B$_{s}\rightarrow \mu^{+}\mu^{-}$;
(*) D0 run IIa and IIb combined result.
\label{tab:exp2}}
\begin{center}
\begin{tabular}{|c|c|c|c|c|c|c|c|}
\hline
Exp. 
& Int. Lum.& \mco{2}{|c|}{B$_{s}$ events} & \mco{2}{|c|}
{B$_{d}$ events} & 
$\mathcal{B}(B_{s} \rightarrow \mu^{+} \mu^{-})$ & 
$\mathcal{B}(B_{d} \rightarrow \mu^{+} \mu^{-})$\\
\cline{3-4} \cline{5-6}
& pb$^{-1}$ & Expected & Obs. & Expected & Obs. & 
$90\%(95\%)$ C.L. & $90\%(95\%)$ C.L. \\
\hline \hline 
CDF & 780 & 1.27$\pm$0.37 & 1 & 2.45$\pm$0.40 & 2 & $<8.0\cdot 10^{-8}$(10) &
$<2.3\cdot 10^{-8}$(3) \\ \hline
D0 & 
\begin{minipage}{2.4cm}{1300 (run IIa)  \\ \hspace*{0.1cm} 500 (run IIb) }
\end{minipage} &
\begin{minipage}{1.7cm}{0.8 \hspace*{0.1cm}$\pm$0.2  \\ 1.5 
\hspace*{0.1cm}$\pm$0.5 }
\end{minipage} &  
\begin{minipage}{1cm}{\hspace*{0.4cm}1 \\ \hspace*{0.4cm}2}
\end{minipage}& -  & - &
$<7.5\cdot 10^{-8}$(9.3)(*)  
& - \\ \hline
\hline
\end{tabular}
\end{center}
\end{table}

CDF and D0 analyzes to reconstruct the B$\rightarrow \mu^{+} \mu^{-} ~h_{s}$
modes are similar; each signal is normalized to the analogous
B$\rightarrow J/\psi ~h_{s}$ mode after a blind cut optimization
and background estimation from the signal sidebands; details of
the procedure can be found elsewhere \cite{CDFD02}.

CDF results for the B$_{u,d}$ modes, obtained with 1 fb$^{-1}$, 
are summarized in table \ref{tab:exp3};
they are in good agreement with those from the B-Factories \cite{BaBel} 
and have similar uncertainty.
\begin{table}[htb]
\caption{CDF result summary for the B$_{u,d}\rightarrow h ~\mu^{+}\mu^{-}$ 
modes;
\label{tab:exp3}}
\begin{center}
\begin{tabular}{|c|c|c|c|}
\hline
Mode & Evts. in the signal region  & Estimated bkg.& BR$\cdot 10^{-6}$ \\
\hline \hline 
B$^{+}\rightarrow \mu^{+}\mu^{-}K^{+}$ & 90 & 45.3$\pm$5.8 & 
0.60$\pm$0.15(stat.)$\pm$0.04(syst.)\\ \hline
B$_{d}\rightarrow \mu^{+}\mu^{-}K^{*0}$ & 35 & 16.5$\pm$3.6 & 
0.82$\pm$0.31(stat.)$\pm$0.10(syst.)\\
\hline
\end{tabular}
\end{center}
\end{table}

CDF and D0 limits for $\mathcal{B}(B_{s}\rightarrow \phi ~\mu^{+}\mu^{-})$ 
are shown in table \ref{tab:exp4}; CDF uses a Bayesian approach to extract
its limit, while D0 confidence interval is constructed within the 
Feldman-Cousins scheme.
\begin{table}[htb]
\caption{Tevatron result summary for B$_{s}\rightarrow \phi ~\mu^{+}\mu^{-}$;
\label{tab:exp4}}
\begin{center}
\begin{tabular}{|c|c|c|c|c|}
\hline
Exp. & Int.Lum. & Observed events & Expected events & BR  \\
\hline \hline 
CDF & 920 pb$^{-1}$ & 11 & 3.5$\pm$1.5 & $<2.4\cdot 10^{-6} (90\%$ C.L.)\\
\hline
D0 & 450 pb$^{-1}$ & 0 & 1.6$\pm$0.6 & $<3.3\cdot 10^{-6} (90\%$ C.L.)\\
\hline
\end{tabular}
\end{center}
\end{table}

\section{Conclusions}

Tevatron is demonstrated to be a good place to study B rare decays offering
different possibilities to constrain more and more New Physics in the
B$_{s}\rightarrow \mu^{+}\mu^{-}$ and $b \rightarrow s ~\mu^{+}\mu^{-}$
modes; moreover, a physics program complementary to the B-Factories can be
exploited in the B$_{s}\rightarrow h^{+}h^{-}$ sector. CDF made the first
observation of the B$_{s}\rightarrow K^{-}\pi^{+}$ mode whose direct CP 
asymmetry appears to be large in agreement with expectation.

Tevatron current values for the limits on $\mathcal{B}
(B_{s}\rightarrow \mu^{+}\mu^{-})$ are entering the 10$^{-8}$ territory; D0
current best limit is the first Tevatron result with 2 fb$^{-1}$
integrated luminosity. Any signal at Tevatron before its run II
program end ($\int L$dt $\sim8$ fb$^{-1}$) will be evidence of New Physics.

CDF and D0 are beginning the exploration of the $b 
\rightarrow s ~\mu^{+}\mu^{-}$ field; upper limits for $\mathcal{B}
(B_{s}\rightarrow \phi ~\mu^{+}\mu^{-})$ obtained with less than
1 fb$^{-1}$ are close to the SM prediction, CDF reported new results
on $\mathcal{B}(B_{u,d}\rightarrow s ~\mu^{+}\mu^{-})$ in general
agreement with the B-Factories; significantly improved results
will come soon.



\section*{References}
\begin{thebibliography}{99}

\bibitem{det} CDF Collaboration, D.Acosta {\it et al},
\Journal{\PRD}{71}{032001}{2005}, \\
D0 Collaboration, V.M.Abazov {\it et al}, \Journal{\NIMA}{565}{463}{2006}.  

\bibitem{Bla} M. Blanke {\it et al.}, \Journal{\JHEP}{610}{003}{2006}

\bibitem{Hur} T. Hurth, submitted to $Nucl. ~Phys.$ B, hep-ph/0612231.

\bibitem{Log} H.E Logan and U. Nierste, \Journal{\NPB}{586}{39}{2000}

\bibitem{Bab} K.S. Babu and C.F. Kolda, \Journal{\PRL}{84}{228}{2000}

\bibitem{SVT} W. Ashmanskas {\it et al}, \Journal{\NIMA}{518}{532}{2004}.

\bibitem{CDF1} CDF Collaboration, A. Abulencia {\it et al},
\Journal{\PRL}{97}{211802}{2006}

\bibitem{Lip} H.J. Lipkin,\Journal{\PLB}{621}{126}{2005}  

\bibitem{HAF} {\it http://www.slac.standford.edu/xorg/hfag}

\bibitem{CDFD02} D0 Collaboration, V.M. Abazov {\it et al.},\Journal{\PRD}{74}{031107(R)}{2006} \\
CDF Collaboration, CDF public note 8543, Nov. 30, 2006 

\bibitem{BaBel} BaBar Collaboration, {\it B. Aubert et al.},\Journal{\PRD}{73}{092001}{2006} \\
Belle Collaboration, K. Abe {\it et. al.}, ICHEP04, Beijing, 16-22 Aug. 2004, hep-ex/0410006.

\end{thebibliography}
\end{document}